\documentclass[journal=jacsat,manuscript=article]{achemso}

\usepackage{achemso}
\usepackage{placeins}
\usepackage{graphics}
\usepackage{amssymb,amsfonts}
\usepackage{mathtools}
\usepackage{physics}
\usepackage{subcaption}
\usepackage{graphicx}
\usepackage[table,dvipsnames]{xcolor}
\usepackage{multirow}
\usepackage{caption}
\usepackage{booktabs}
\usepackage{colortbl}
\usepackage{amsmath}
\usepackage{amsopn}
\usepackage{bm}
\usepackage{color}
\usepackage{array}
\usepackage{lscape}
\usepackage{mciteplus}
\usepackage[version=3]{mhchem}
\usepackage[normalem]{ulem}
\usepackage{listings}
\usepackage{enumerate}
\usepackage{lmodern}

\newcommand{\onlinecite}[1]{\hspace{-1 ex} \nocite{#1}\citenum{#1}} 

\author{Malte Hellmann}
\email{mhellma@uni-mainz.de}
\affiliation[mainz]{Department Chemie, Johannes Gutenberg-Universit{\"a}t Mainz\\Duesbergweg 10--14, 55128 Mainz, Germany}
\author{J{\"u}rgen Gauss}
\email{gauss@uni-mainz.de}
\affiliation[mainz]{Department Chemie, Johannes Gutenberg-Universit{\"a}t Mainz\\Duesbergweg 10--14, 55128 Mainz, Germany}

\title{The Direct-Product Decomposition Approach for Symmetry Exploitation
in Many-Body Methods in Case of Non-Abelian Point Groups}

\begin{document}

\begin{abstract}

\noindent We demonstrate for the specific case of $C_{3v}$ how the direct-product decomposition scheme for the treatment of symmetry in coupled-cluster (CC) calculations can be extended to non-Abelian point groups. We show that for the two-electron integrals and CC amplitudes a block structure can be obtained by resolving the reducible products of two irreducible representations
into their irreducible representations. To deal with the necessary resorts of the ordering of the two-electron integrals and amplitudes, spin-adaptation, and the ${\cal O}(M^5)$ contractions (with $M$ as the number of basis functions) of a CC calculation, we suggest a strategy that uses both the reduced and non-reduced representation of the corresponding quantities and switches back and forth between them. While the reduced representations are the ones used in the ${\cal O}(M^6)$ contractions, the other steps are better carried out in the non-reduced representation. Our pilot implementation of the CC singles and doubles method confirms in test calculations for NH$_3$ and PH$_3$ using different basis sets that significant savings (of more than 20 compared to treatments without symmetry and about 5 compared to treatments using $C_s$ symmetry) are possible and suggest
that the exploitation of non-Abelian symmetry would render CC computations on large highly symmetric molecules possible.
\end{abstract}

\section{Introduction}
Symmetry is a very important concept in all areas of natural science.\cite{Mainzer88} For chemistry, symmetry is useful for the explanation of chemical bonding (within molecular-orbital theory\cite{Bishop73} or ligand-field theory\cite{Ballhausen62}), the interpretation of spectra via selection rules,\cite{Bunker06} the explanation of chemical reactivity via the Woodward-Hoffmann rules,\cite{Woodward69} etc. 
However, symmetry can be also exploited to facilitate and speed up computational simulations of all kinds. The main point is then that the use of symmetry can lead to a reduction in the computational cost and thus make simulations faster or even feasible in the first place. This aspect of symmetry has turned out particularly useful in quantum chemistry, where computations often are very costly in terms of both computing time and memory requirement. Symmetry has been thus considered straight from the beginning\cite{Roothaan51} and since then many publications have dealt with the use of symmetry in quantum-chemical computations.\cite{Davidson75,Dupuis77,Dupuis78,Carsky84,Taylor85,Taylor86,Ahlrichs89,Haeser89,Stanton91a,Gauss91a,Haeser91b, Haeser91c,Almlof97,Kollwitz98,Tayloresqc,Nottoli23,Melega26} 

The key in the exploitation of symmetry in quantum chemistry lies in the fact that the solutions of the electronic Schr\"odinger equation (and also of the Hartree-Fock (HF) equations, etc.) are required to have symmetry properties imposed by the underlying symmetry of the molecule. The solutions of the electronic Schr\"odinger equation, etc. thus have to transform as one of the irreducible representations of the corresponding point group of the atom or molecule. Symmetry considerations furthermore lead to selection rules that render certain quantities (e.g., one- and two-electron integrals, amplitudes in perturbation and coupled-cluster (CC) theory, etc.) to vanish which can be exploited in quantum-chemical computations. The literature documents many examples for the usage of symmetry in quantum-chemical computations and impressive savings due to the exploitation of symmetry have been reported.

To be more specific, symmetry can be exploited in quantum-chemical computations in (a) the integral evaluation\cite{Davidson75,Dupuis77, Dupuis78,Taylor86,Almlof97} and the self-consistent-field (SCF) step,\cite{Dupuis77,Ahlrichs89,Almlof97} (b) the transformation of the integrals from the atomic-orbital (AO) to the molecular-orbital (MO) representation,\cite{Carsky84,Haeser91b} and (c) within the electron-correlation treatment.\cite{Stanton91a,Gauss91a,Kollwitz98,Nottoli23,Melega26} In the first of the three tasks the challenge is to exploit symmetry in the evaluation and handling of quantities that are not necessarily symmetry-adapted (i.e., that do not transform as one of the irreducible representations of the point group of the molecule). This is due to the fact that the underlying AOs are not necessarily symmetry-adapted. On the other side,
electron-correlation treatments are most often formulated in terms of MOs and all involved quantities (integrals and amplitudes in the case of CC computations, for example) are symmetry-adapted. For the treatment of the latter in CC computations, Stanton {\it et al.}\cite{Stanton91a} devised an elegant and efficient scheme for the exploitation of symmetry in case of Abelian point groups. Their direct-product decomposition (DPD) scheme enables a symmetry blocking of all quantities and can be implemented without a significant overhead. However, the scheme in its original version is restricted to Abelian point groups, as it relies on the fact that the direct product of two irreducible representations is another irreducible representation, something which does no longer hold for non-Abelian point groups. Attempts to extend the DPD scheme to non-Abelian point groups can be traced back to the 90ies (unpublished work by Stanton and Gauss), but those attempts were never pushed to a working code.

In this paper, we resume this issue and reinvestigate the possibility to fully exploit non-Abelian point-group symmetry in electron-correlated calculations. We will report a pilot implementation of the CC singles and doubles (CCSD) method\cite{Purvis82} and will demonstrate that the exploitation of non-Abelian point-group symmetry can lead to significant savings and, thus, is worthwhile to be considered.

We start our discussion in the next section (section 2) by briefly reviewing the DPD scheme for exploiting symmetry in case of Abelian point groups. This is followed by a description of our proposed extension of the DPD scheme to the special case of $C_{3v}$ symmetry, i.e., one of the simplest non-Abelian point groups. Details about our pilot program are given in section 3, and in the result section (section 4) we report and discuss the operation counts for CCSD computations on NH$_3$ and PH$_3$ using different basis sets. We conclude with a summary and an outlook (section 5) and there in particular discuss the perspectives for a general implementation for arbitrary non-Abelian point groups.

\section{Symmetry in Quantum-Chemical Calculations}

\subsection{General Concepts}

Symmetry selection rules generally can be used to decide
which quantities need to be computed, stored, and processed. For example, a quantity 
\begin{eqnarray}
A_{pqr \dots} = \int a_p a_q a_r ... d \tau,
\end{eqnarray}
where the functions $a_p, a_q, a_r, \dots$ can be classified according to the irreducible representations of the given point group and integration is over all variables (collectively referred to as $\tau$), can only have a nonzero value if the direct product of the irreducible representations of the functions $a_p, a_q, a_r, \dots$ contains the totally symmetric representation.

In quantum chemistry, the molecular orbitals (MOs) $\phi_p$ are often obtained as solutions of the canonical HF equations and accordingly transform as the irreducible representations of the point group of the considered molecule. It is of advantage to order them, occupied orbitals (indices $i, j, k, \dots$) and virtual orbitals (indices $a, b, c, \dots$) separately, so that those of the same irreducible representation are grouped together, i.e.,
\begin{eqnarray}
\{ \phi_1^{\mathrm irrep1}, \phi_2^{\mathrm irrep1} \dots \phi_{n^{\mathrm irrep1}}^{\mathrm irrep1},\phi_1^{\mathrm irrep2} \phi_2^{\mathrm irrep2} \dots \}
\end{eqnarray}
with $n^{\mathrm irrep1}, n^{\mathrm irrep2}, \dots$ as the number of occupied/virtual orbitals per irreducible representation.
Using this ordering, two-dimensional objects such as, for example, the one-electron integrals (here and in the following we always assume that the operator that appears in the integrals transforms as the totally symmetric representation)
acquire a block structure in the following referred to as block matrix(see Scheme \ref{scheme1}).
\begin{scheme}
  \includegraphics[width=5cm, height=!]{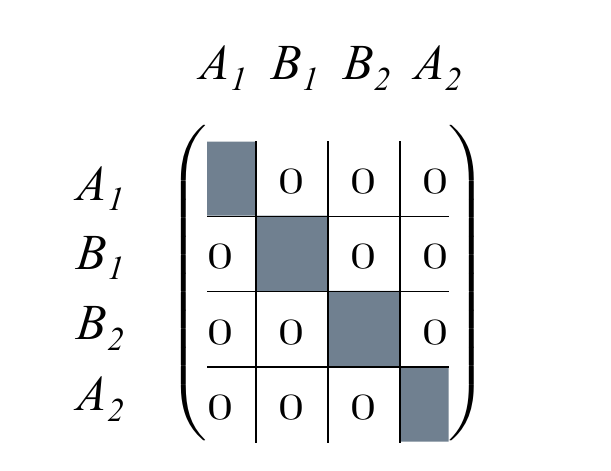}
  \caption{Block structure of a two-dimensional matrix in case of $C_{2v}$ symmetry with four irreducible representations.}
  \label{scheme1}
\end{scheme}
This block structure turns out to be useful also for matrix products, as the product of two block matrices is easily obtained by just multiplying the individual blocks with each other (see Scheme ~\ref{scheme2}).
\begin{scheme}
  \includegraphics[width=10cm, height=!]{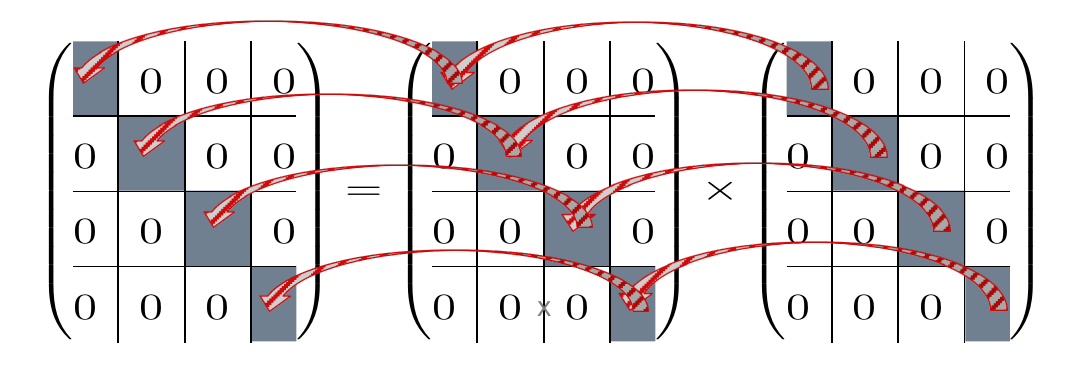}
  \caption{Block-by-block multiplication of two block matrices.}
  \label{scheme2}
\end{scheme}

For higher-dimensional quantities such as the two-electron integrals, amplitudes from CC computations, etc., the discussion is more involved, as a convenient block structure is not automatically obtained like in the case of the two-dimensional matrices.

\subsection{Direct-Product Decomposition in the Case of Abelian Point Groups}

In the case of Abelian point groups, all irreducible representations are 
one-dimensional and the direct product of two irreducible representations is again a one-dimensional irreducible representation. This property of Abelian point groups can be, as shown in Ref.~\onlinecite{Stanton91a}, used to establish a block structure even for higher-dimensional matrices. Focusing on four-dimensional matrices (two-electron integrals, double-excitation amplitudes in CC computations, etc.) a block structure can be obtained by considering those matrices as two-dimensional objects with super indices formed from
the original indices. The irreducible representation of the super index is then just the direct product of the indices that define the super index. For a two-electron integral (and in the same way for the double-excitation amplitudes $t_{ij}^{ab}$)
\begin{eqnarray}
\langle p q | r s \rangle = \int d^3 r_1 \int d^3 r_2\  \varphi_p ({\bf r}_1) \varphi_q({\bf r}_2) \frac{1}{r_{12}} \varphi_r({\bf r}_1)\varphi_s({\bf r}_2)
\end{eqnarray}
a two-dimensional quantity $I_{(pq);(rs)}$ can be defined with  super indices formed from the indices $p$ and $q$ as well from $r$ and $s$, respectively.\footnote{Note that this is just one way to define a two-dimensional quantity, as other ordering such as $I(pr,qs)$, $I(ps,qr)$ are also possible. The actual choice depends on the targeted contraction and, for example, is different for the ladder and ring terms in CC treatments.} For $I_{(pq);(rs)}$
a block structure is obtained in the same way as for simple two-dimensional matrices (see Scheme~\ref{scheme3}). 
\begin{scheme}
  \includegraphics[width=10cm, height=!]{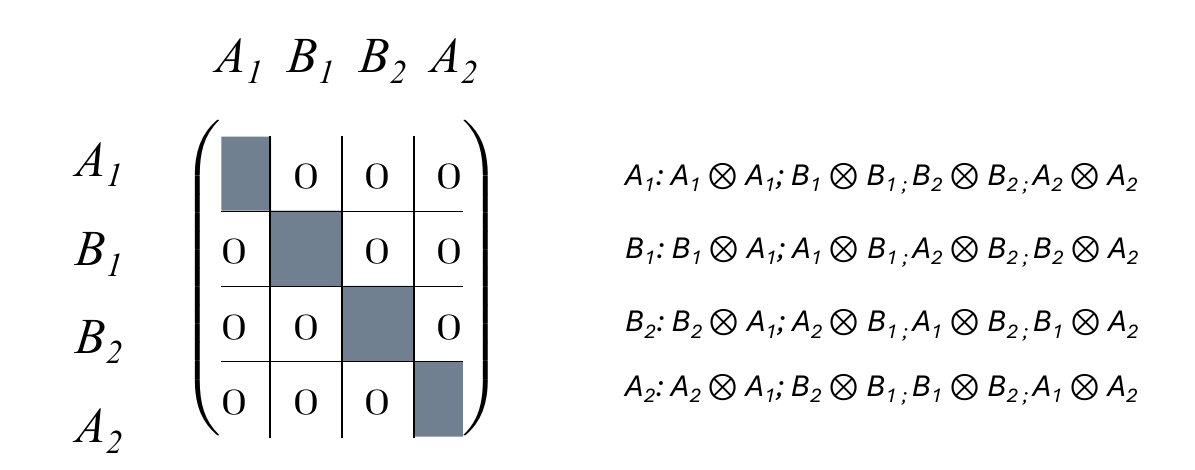}
  \caption{Block structure of a four-dimensional matrix in case of $C_{2v}$ symmetry with four irreducible representations. The information on the right gives all products of two irreducible representations that correspond to a super index of given symmetry.}
  \label{scheme3}
\end{scheme}

\begin{figure}
    \centering
    \includegraphics[width=10cm, height=!]{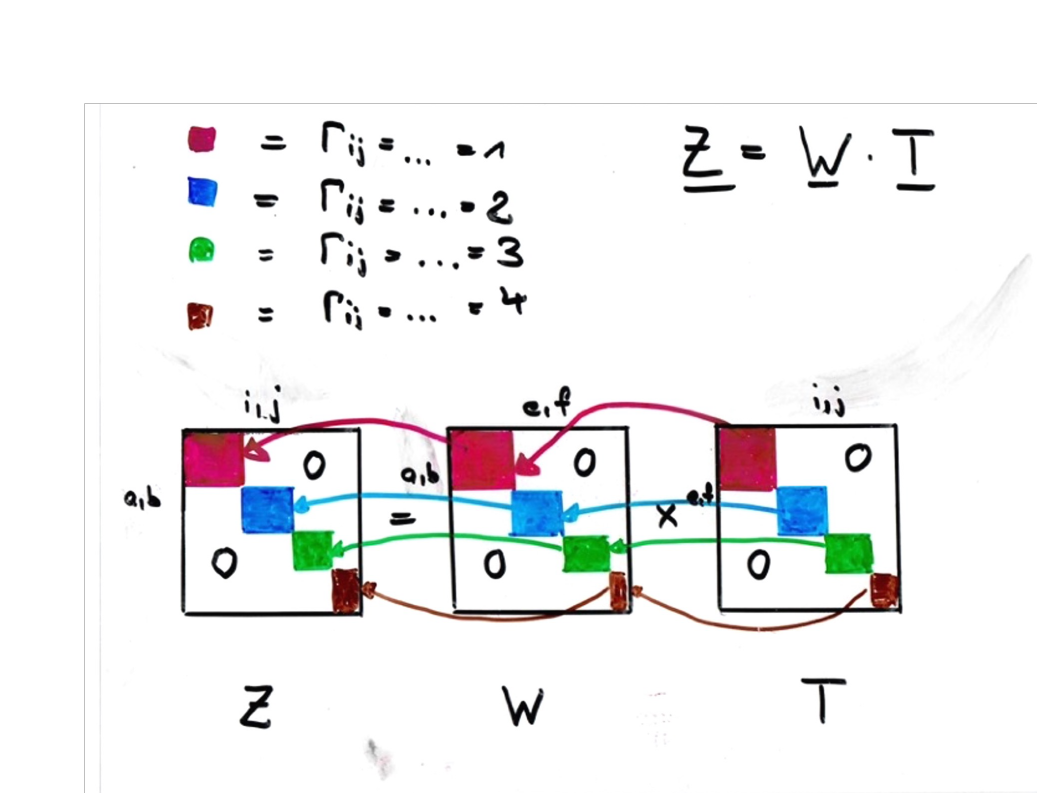}
    \caption{Slide from one of the author's talk (J.G. together with John F. Stanton, John D. Watts, and Rodney J. Bartlett) at the 31rst Sanibel Symposium 1991 in St. Augustine, (Florida, USA) explaining the block-by-block multiplication in the PPL terms of CCSD within the DPD scheme.}
    \label{figure1}
\end{figure}

As this procedure results in a decomposition of the direct product of four irreducible representations in the direct product of two direct products of two irreducible representations, it has been coined direct-product decomposition (DPD) approach. The resulting block structure for four and higher-dimensional quantities significantly simplifies the exploitation of symmetry in electron-correlated computations, as a contraction like the particle particle ladder (PPL) term in CC computations
can now be performed in a block-wise manner. Figure~\ref{figure1} 
explains the procedure. The resulting savings are in the ideal case of the square of the order of the group and actual computations indeed show savings (in the operation count) close to the optimal value.\cite{Stanton91a} The actual savings in the computation times, however, are somewhat lower, as the symmetry blocking reduces the size of the matrices and thus slows down (for example, when optimized BLAS routines are used for the multiplication) the speed of the calculations. Even more importantly, the DPD scheme enables the exploitation of symmetry with a minimum of overhead, as no extensive symmetry checks are needed after the used quantities have been ``symmetry blocked". 

\subsection{Extension of the Direct-Product Decomposition Scheme to $C_{3v}$ Symmetry}

The challenge when dealing with non-Abelian point group symmetries is that the direct product of two irreducible representations is generally no longer given by one irreducible representation but rather as a direct sum of several irreducible representations. For that reason a simple DPD as in the Abelian case does not lead to a useful
block structure for higher-dimensional matrices. We will discuss this issue in the following for the specific case of $C_{3v}$. This point group contains 6 symmetry operations, namely $E$, $C_3$, $C_3^2$, $\sigma_{v1}$, $\sigma_{v2}$, and $\sigma_{v3}$. It has 3 irreducible representations ($A_1$, $A_2$, and $E$), i.e., two one-dimensional ones and a two-dimensional one. Its character table is given by  
\begin{center}
\begin{tabular}{ c | c c c }
 & $E$ & 2 $C_3$ & 3$\sigma_v$ \\ 
 \hline
 $A_1$ & 1 & 1 & 1\\  
 $A_2$ & 1 & 1  & -1  \\
 $E$ & 2 & -1 & 0    
\end{tabular}
\end{center}
and, thus, the direct product of $E$ with $E$ is
\begin{eqnarray}
E \otimes  E = A_1 \oplus A_2 \oplus E.
\end{eqnarray}
We will use in the following a particular choice for the $E$ representation in the way that one component transforms in $C_s$ (the largest Abelian subgroup of $C_{3v}$) as $A^\prime$ and the other as $A^{\prime\prime}$. Accordingly, we denote these components as $E(A^\prime$) and $E(A^{\prime\prime}$). Note that the different components of higher-dimensional irreducible representations always can be chosen in such a manner and that in this way the discussion of $C_{3v}$ suffices to illustrate our suggested adaption of the DPD scheme to non-Abelian symmetries. However, our discussion does not apply to complex Abelian point groups (e.g., $C_3$) which in a real representation also have higher-dimensional irreducible representations.

A straightforward application of the DPD scheme would yield for four-index quantities such as the two-electron integrals to the block structure shown in Scheme~\ref{scheme4}
\begin{scheme}
  \includegraphics[width=12cm, height=!]{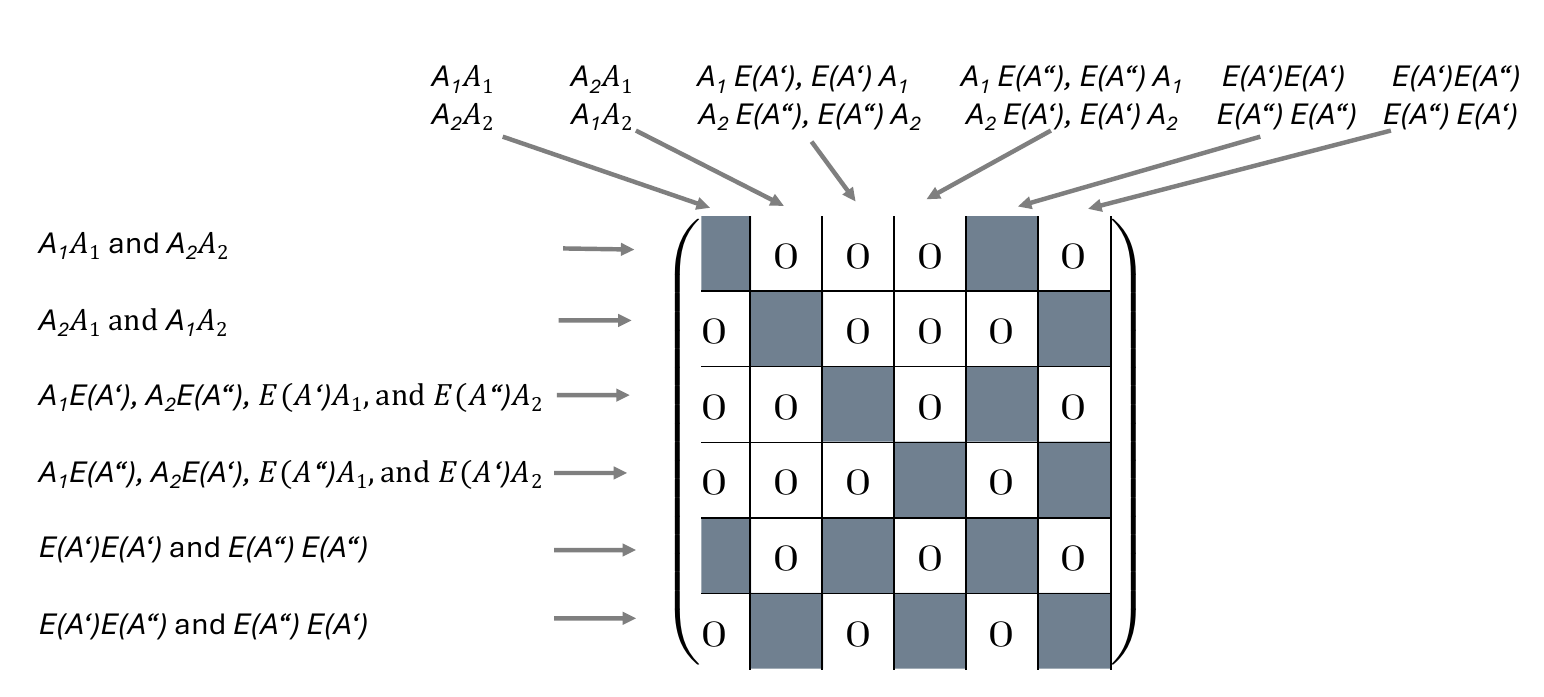}
  \caption{Block structure of a four-dimensional matrix in case of $C_{3v}$ symmetry without reduction of the $E \otimes E$ blocks.}
  \label{scheme4}
\end{scheme}
which apparently contains off-diagonal blocks and thus is not particularly useful for performing contractions in electron-correlated computations. Another problem of this block structure is the redundancy of information, i.e., integrals with the same value appear twice which is caused by the two-dimensional $E$ representation. A diagonal block structure, however, is recovered if the direct product $E\otimes E$ is reduced, which can be done using projection operators\cite{Tayloresqc} and leads to
 \begin{eqnarray}
 E \otimes E (A_1) & =& \frac{1}{\sqrt{2}} E(A^{\prime}) \cdot E(A^{\prime}) + \frac{1}{\sqrt{2}} E(A^{\prime\prime}) \cdot E(A^{\prime\prime}) \\
E \otimes E (A_2) &=& \frac{1}{\sqrt{2}} E(A^{\prime}) \cdot E(A^{\prime\prime}) - \frac{1}{\sqrt{2}} E(A^{\prime\prime}) \cdot E(A^{\prime})   \\
E \otimes E (E, {\rm first\ component}) &=& \frac{1}{\sqrt{2}} E(A^{\prime}) \cdot E(A^{\prime}) - \frac{1}{\sqrt{2}} E(A^{\prime\prime}) \cdot E(A^{\prime\prime})  \\
E \otimes E (E, {\rm second\ component}) &=& - \frac{1}{\sqrt{2}} E(A^{\prime}) \cdot E(A^{\prime\prime}) - \frac{1}{\sqrt{2}} E(A^{\prime\prime}) \cdot E(A^{\prime}).  
\end{eqnarray}
With the reduction carried out, a diagonal block structure as shown in Scheme~\ref{scheme5} can be recovered
and the redundancy in the $E$ blocks can be eliminated by simply dropping the second $E$ block.
\begin{scheme}
  \includegraphics[width=10cm, height=!]{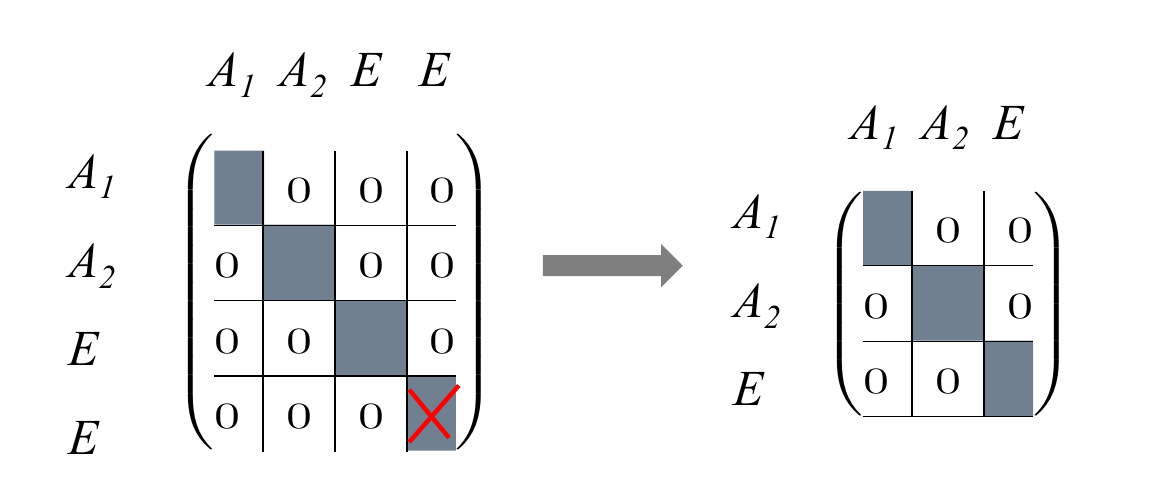}
  \caption{Block structure of a four-dimensional matrix in case of $C_{3v}$ symmetry after reduction  with three irreducible representations; the block that has been crossed out is redundant and can be skipped. The two $E$ blocks in the first matrix corresponds to the blocks containing the first and second component of the $E$ representation, respectively.}
  \label{scheme5}
\end{scheme}

\noindent
The reduced DPD representation is obtained by first sorting the non-reduced quantities
(in case of the two-electron integrals the integrals obtained in an integral transformation using the largest Abelian subgroup, i.e., in our case $C_s$ symmetry)
in such a manner that the $I(E(A^\prime)  E(A^\prime), \\ E(A^\prime) E(A^\prime))$ elements are in the $A_1$ block, the $I(E(A^\prime) E(A^{\prime\prime}),E(A^\prime) E(A^{\prime\prime}))$ elements in the $A_2$ block and the $I(E(A^{\prime\prime}) E(A^{\prime\prime}), E(A^{\prime\prime}) E(A^{\prime\prime}))$ elements in the $E$ block (see Scheme \ref{scheme6}). Note that we already at this point do not need all elements of $(E E, E E)$ type and in this way eliminate redundancies. Redundancies are also avoided by just storing those of $I(E(A^\prime)A_1,E(A^\prime)A_1)$ and $I(E(A^\prime) A_2, E(A^\prime) A_2)$ type and not the redundant $I(E(A^{\prime\prime})A_1, E(A^{\prime\prime})\\ A_1)$ and $I(E(A^{\prime\prime}) A_2, E(A^{\prime\prime}) A_2)$ elements.
\begin{scheme}
  \includegraphics[width=10cm, height=!]{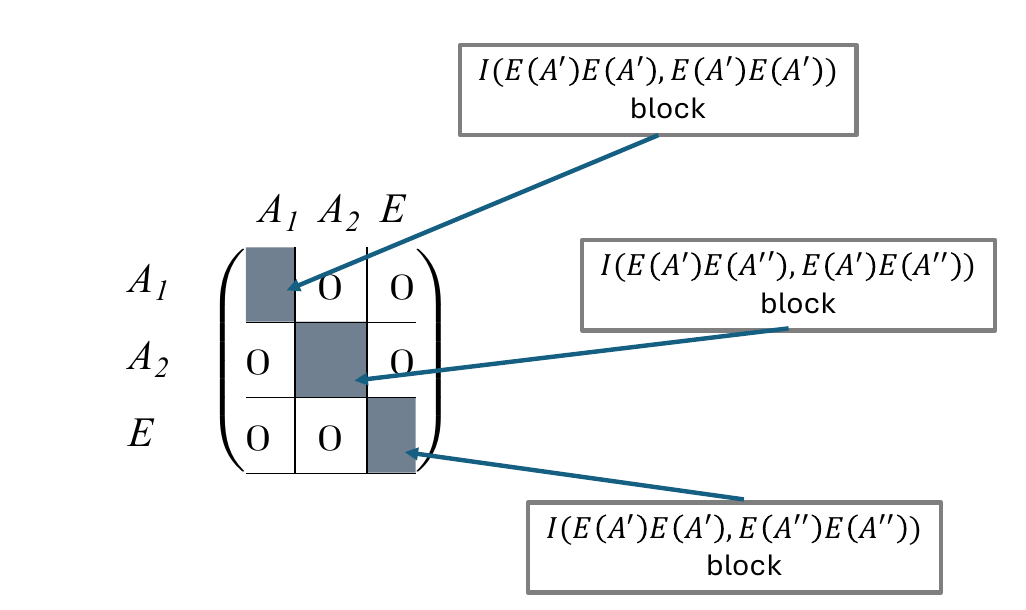}
  \caption{Sorting of a four-dimensional quantity (e.g, two-electron integrals or double-excitation amplitudes) in a non-reduced block matrix. Note that the designation of the blocks are only meant to indicate the subblock in which the corresponding non-reduced elements are incorporated }
  \label{scheme6}
\end{scheme}

From this non-reduced representation the reduced one is easily obtained via
\begin{eqnarray}
(E \times E) (A_1)\cdot(E \times E) (A_1) &=& (E(A') \times E(A'))\cdot(E(A') \times E(A')) \nonumber \\ && +(E(A') \times E(A'))\cdot (E(A'') \times E(A'')) 
\label{reduce1}
\\
(E \times E) (E)\cdot(E \times E) (E) &=& (E(A') \times E(A')) \cdot(E(A') \times E(A')) \nonumber \\&&-(E(A') \times E(A'))\cdot (E(A'') \times E(A''))\\
\label{reduce2}
(E \times E) (A_2)\cdot(E \times E) (A_2) &=& 2 (E(A') \times E(A''))\cdot (E(A'') \times E(A'))\nonumber \\ &&-E(A') \times E(A'))\cdot (E(A') \times E(A'))
\nonumber \\ && +(E(A') \times E(A')) \cdot (E(A'') \times E(A'')). \ 
\label{reduce3}
\end{eqnarray}
Note that the given expressions account for the fact that reduction is necessary for both the right and left side of the four-index quantity.
In addition, reduction requires that all elements of the four-index quantity that possess either on the right or left side $E \times E$ symmetry (except those with all four orbitals belonging to the $E$ representation) are scaled by $\sqrt{2}$.

Using the block structure of the reduced matrices 
contractions like the one required for the PPL term can be carried out in the same manner as for the Abelian case, i.e., block by block (see Scheme~\ref{scheme7}).
\begin{scheme}
  \includegraphics[width=10cm, height=!]{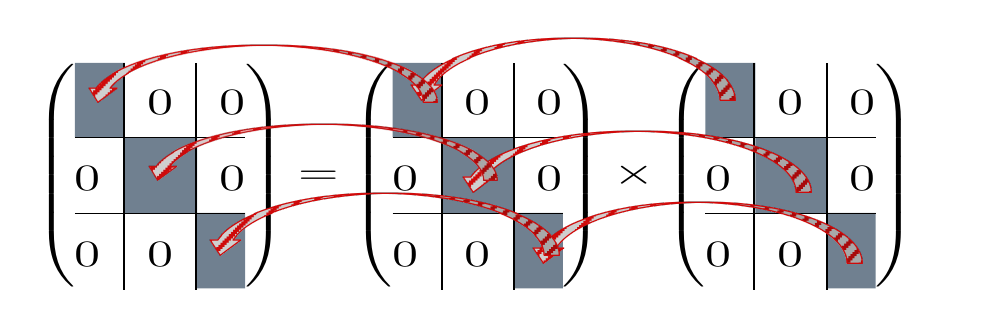}
  \caption{Block-by-block multiplication of two reduced four-dimensional block matrices as required for the ${\cal O}(M^6)$ terms in MP3, CCD, and CCSD computations.}
  \label{scheme7}
\end{scheme}

The discussion so far enables an efficient handling of non-Abelian symmetry of the ${\cal O}(M^6)$ terms in third-order M{\o}ller-Plesset (MP3) perturbation theory,\cite{Cremer00} CC doubles (CCD)\cite{Taylor76,Pople78,Bartlett78} as well as CCSD
computations\cite{Purvis82} provided the required two-electron integrals and double-excitation amplitudes are ordered properly. For actual CCD/CCSD computations,  however, additional issues need to be discussed, as there are (a) necessary resorting steps required for the four-index quantities (for different contractions involving the same four-index quantity different orderings of its elements are needed and those are subject to different DPDs), (b) formation of spin-adapted quantities needed for an efficient closed-shell treatment, and (c) handling of the additional ${\cal O}(M^5)$ terms in CCD and CCSD calculations. 
These issues are trivially dealt with in the case of Abelian point-group symmetry, but require some additional thoughts in case of non-Abelian symmetry. 

\subsection{Non-Reduced and Reduced Representations}
\begin{figure}
    \centering
    \includegraphics[width=10cm, height=!]{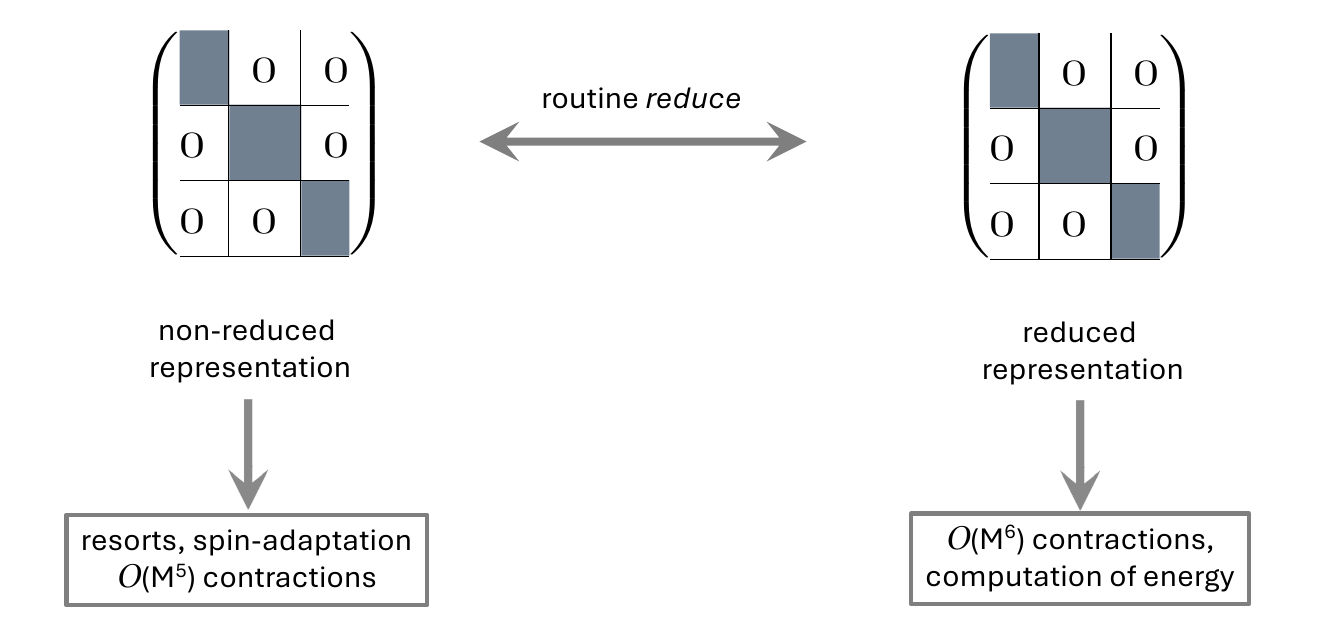}
    \caption{Simultaneous usage of non-reduced and reduced representations in CC computations exploiting non-Abelian symmetry.}
    \label{figure2}
\end{figure}

The key idea to resolve the raised issues is to use both a reduced and a non-reduced representation for the four-index quantities of interest (see Figure~\ref{figure2}). The reason for doing this is simply that resorts are straightforward in the non-reduced representation (as all entries there represent quantities with four specific indices assigned unlike for the reduced representation where the entries are actually linear combinations of such quantities), while the ${\cal O}(M^6)$  
contractions are efficiently done using the reduced representation. However, it is not necessary to keep simultaneously both representations in memory or on disk, as it is very simple to switch back and forth (see Eqs.~(\ref{reduce1}) to (\ref{reduce2})). A key element of our implementation is therefore a routine (in our implementation called $reduce$) that does exactly that as shown in Figure~\ref{figure2}.

\subsection{Resort of Reduced Representations of Four-Index Quantities}
For the resort of a four-index quantity we suggest to first revert back to the non-reduced representation, then do the resort, and finally reduce the resorted quantity, as shown in Figure~\ref{figure2}.
A complication arises here due to the fact that the non-reduced representation does not involve all elements of the four-dimensional quantity (the ``missing integral/amplitude'' issue) that are needed for the sort. For example, for a resort (12,34) $\rightarrow$ (14,23), one has in the original non-reduced representation all quantities of type $E(A') E(A') E(A') E(A')$, $E(A') E(A') E(A'') E(A'')$,
and $E(A') E(A'') E(A') E(A'')$, but misses those of $E(A') E(A'') E(A'')$ $ E(A')$.
However, the missing quantity is given by
\begin{eqnarray}
I(E(A')E(A''), E(A'')E(A'))  &=& I(E(A')E(A'), E(A')E(A')) \nonumber \\ &&-I(E(A')E(A'), E(A'')E(A''))\nonumber\\ &&-I(E(A')E(A''), E(A')E(A''))
\label{missing}
\end{eqnarray}
and, thus, a resort of the non-reduced quantity is possible, though the corresponding routine is more involved than in the Abelian case.
Two examples of resorts (i.e., (1234) $\rightarrow$ (1324), where no problem  due to ``missing integrals/amplitudes'' appears, and (1234) $\rightarrow$ (1423), where the ``missing integral/amplitude'' issue has to be dealt with) are explained in Scheme~\ref{scheme8}
\begin{scheme}
  \hspace{1.7cm}\includegraphics[width=12cm, height=!]{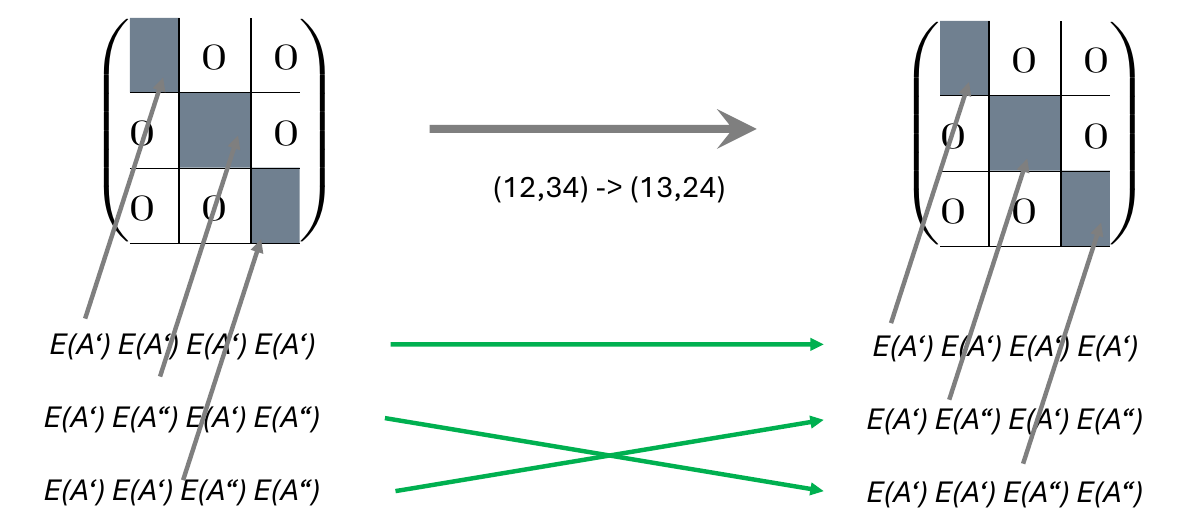}
  \vspace{0.5cm}\includegraphics[width=15cm, height=!]{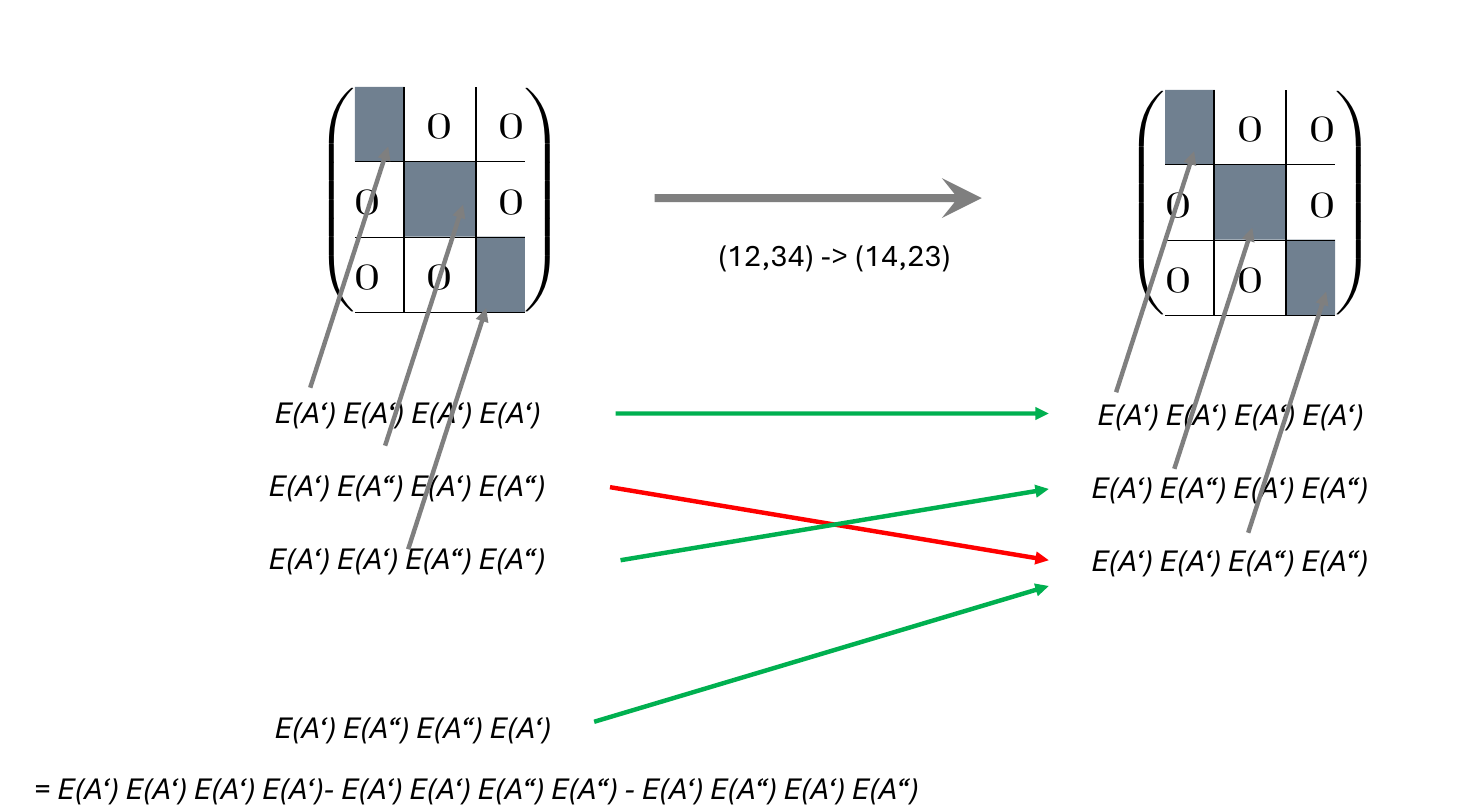}
  \caption{(1234) $\rightarrow$ (1324) and (1234) $\rightarrow$ (1423) resorts for non-reduced block matrices. The red arrow shows the part of the resort affected by the ``missing integral/amplitude'' issue.}
  \label{scheme8}
\end{scheme}

\subsection{Spin-Adaptation of Reduced Representations of Four-Index Quantities}
The efficient formulation of MP and CC models in case of a closed-shell reference (i.e., a restricted HF reference determinant) involves so-called spin-adapted quantities. For example, in case of the double-excitation amplitudes they are given by
\begin{eqnarray}
{\widetilde {t_{ij}^{ab}}} = 2 t_{ij}^{ab} - t_{ij}^{ba}
\label{sat}
\end{eqnarray}
However, the reduced amplitudes cannot be ``spin-adapted" using Eq.~(\ref{sat}), as the reduced spin-adapted amplitudes are actually defined by applying the reduction to the non-reduced spin-adapted amplitudes. Again, the solution is to switch back to the non-reduced representation, spin-adapt, and to obtain the reduced spin-adapted quantity by reduction of the non-reduced spin-adapted quantity. The corresponding routine is straightforward to set up, but again more involved than in the Abelian case, as we again have to deal with the ``missing integral/amplitude" issue.
In order to obtain the spin-adapted $\widetilde{I}(E(A') E(A''), E(A') E(A''))$ elements, we need the $I(E(A') E(A''), E(A'') E(A'))$ elements that are not present in the corresponding block matrix, but the required missing elements can be again reconstructed using Eq.~(\ref{missing}).

\subsection{{\cal O}($M^5$) Contractions in CCD and CCSD Computations}
Besides the time-determining ${\cal O}(M^6)$ terms, CCD and CCSD computations involve additional ${\cal O}(M^5)$ terms. There are two types of them, one which contracts two four-index quantities into a two-index quantity
\begin{eqnarray}
I_1(p,q) = \sum_{r,s,t} I_2(rs,tp) I_3(rs,tq)
\label{contraction1}
\end{eqnarray}
and one that contracts a four-index quantity with a two-index quantity to yield a four-index quantity
\begin{eqnarray}
I_1(pq,rs) = \sum_t I_2(pq,rt) I_3(t,s). 
\label{contraction2}
\end{eqnarray}
While both contractions in principle require new DPDs, where the four irreducible representations of the four-index quantities are decomposed into a product of three irreducible representations and a remaining one, it is in the Abelian case more convenient to perform these contractions using the given DPDs. The same is also true in the case of non-Abelian point-group symmetry except that these contractions then need to be carried out using the corresponding non-reduced representations. The contraction in Eq.~(\ref{contraction2}) is straightforward to carry out in the non-reduced representation (thereby assuming that the two-index quantity is given as a two-dimensional block matrix), while for the first contraction, i.e., the one of Eq.~(\ref{contraction1}), one has to deal again with the ``missing integral/amplitude'' issue. In addition, one has to account for the two-dimensionality of the $E$ representation by including in some terms an additional factor of 2 and 4, respectively.

With this, the required tools for handling non-Abelian point-group symmetry have been identified and set up and will make it possible to perform MP3, CCD, and CCSD computations with exploitation of non-Abelian point-group symmetry. Our implementation is described in the next section, before we demonstrate the potential savings that result from the exploitation of non-Abelian point group symmetry (in comparison to just Abelian point-group symmetry).

\section{Implementation and Computational Details}

We implemented our proposed strategy for exploiting non-Abelian point-group symmetry in a pilot program termed ''Quantum Chemistry with Exploitation of Non-Abelian Symmetry'' ({\sc QUENA})\cite{quena} for performing MP3, CCD, and CCSD computations for closed-shell molecules 
with $C_{3v}$ symmetry. The CCSD equations used in this implementation are modified versions of those reported in Refs.~\onlinecite{Scuseria87a} and \onlinecite{Stanton91a} and exploit spin-symmetry (apart from the use of symmetric/antisymmetric algorithm\cite{Saebo87,Scuseria88c} in the ladder terms), thus ensuring (apart from a factor of two in the ladder terms due to the symmetric/antisymmetric algorithm)  minimal cost. For completeness as well as for the discussion in the next section, we summarize in the following our equations:

\noindent a) energy
\begin{eqnarray}
E =  \sum_{m,n}\sum_{e,f} \langle m n | e f\rangle\ \widetilde{\tau_{m n}^{e f}}
\end{eqnarray}
b) singles equations
\begin{eqnarray}
D_i^a\ t_i^a& = &
\sum_e \left\{{\cal F}_{ae} 
+ \frac{1}{2} \sum_m {\cal F}_{me}\ t_m^a\right\}\ t_i^e- \sum_m \left\{{\cal F}_{mi} - \frac{1}{2} \sum_e {\cal F}_{me}\ t_i^e\right\}\ t_{m}^a + 
\sum_{e} \sum_m {\cal F}_{me}\ \widetilde{t_{i m}^{a e}}
\nonumber \\ &&+ \sum_e \sum_m  \widetilde  {\langle a m | i e\rangle}\
t_{m}^e
-
\sum_{m,n} \sum_e \langle m n | i e\rangle\ \widetilde {t_{m  n}^{a  e}}
+  \sum_{m} \sum_{e,f} \langle a m | e f\rangle\  \widetilde {t_{ i m}^{e f}} 
\end{eqnarray}
c) doubles equations
\begin{eqnarray}
D_{ij}^{ab}\ t_{i j}^{a b}&=& \langle a b| i  j\rangle +  P_{+}(ai,bj)\ \sum_e\ {\cal F}_{ae}\ t_{i j}^{e b}  -  P_{+}(ai,bj)\ \sum_m {\cal F}_{mi}\ t_{m j}^{a b} \nonumber \\ &&
+  \sum_{m,n} {\cal W}_{m n i j}\  \tau_{m n}^{a b} + \sum_{e,f} {\cal W}_{abef}\ \tau_{ij}^{ef} \nonumber \\ &&
+  \frac{1}{2} P_{+}(ia,jb)\ \sum_m \sum_e \ \left[\widetilde{{\cal W}_{m b e j}}\ \widetilde {t_{i m}^{a e}} \ + \ \frac{1}{2} \ {\cal W}_{m b j e} \ t_{m i}^{a e}\ + {\cal W}_{mbie}\ t_{m  j}^{a  e} - \langle mb|ej \rangle \  t_i ^e\ t_m ^a \right.\nonumber \\ && \left. + \langle mb|ie \rangle\ t_j ^e\ t_m ^a \right]
\nonumber \\ &&
 -  P_+(ia,jb)\ \sum_m\langle m  b| i j \rangle\ t_{m}^{a }   + P_+(ia,jb)\  \sum_e \langle a  b | e j \rangle\  t_{i}^{e}.
\end{eqnarray}
In the given equations
the two-electron integrals, in Dirac notation, are denoted by $\langle p  q |r s\rangle$ and
the single- and double-excitation amplitudes are given by $t_i^a$ and $t_{i j}^{a b}$, respectively. The double-excitation amplitudes here correspond to the $\alpha \beta \alpha \beta$ amplitudes within a spin-orbital formulation.\cite{Stanton91a}
In addition, the so-called $\tau$ amplitudes 
\begin{eqnarray}
\tau_{ij}^{ab} = t_{ij}^{ab} + t_i^a\  t_j^b,
\end{eqnarray}
are used as well as 
the corresponding spin-adapted $t$ and $\tau$ amplitudes defined by
\begin{eqnarray}
\widetilde {t_{ij}^{ab}} = 2 t_{i j}^{a b} - t_{ij}^{b a},
\end{eqnarray}
and
\begin{eqnarray}
\widetilde {\tau_{ij}^{ab}} = 2 \tau_{i j}^{a  b} - \tau_{i j}^{b  a}.
\end{eqnarray}
Spin-adapted two-electron integrals are denoted in the given equations by
\begin{eqnarray}
\widetilde {\langle pq |rs\rangle} = 2 \langle pq |rs\rangle - \langle pq |sr\rangle
\end{eqnarray}
and the orbital-energy denominators are given by
\begin{eqnarray}
D_i^a = \frac{1}{\varepsilon_i - \varepsilon_a},
\end{eqnarray}
and
\begin{eqnarray}
D_{ij}^{ab} = \frac{1}{\varepsilon_i + \varepsilon_j - \varepsilon_a - \varepsilon_b}
\end{eqnarray}
with orbital energies $\varepsilon_p$, respectively. Furthermore, the intermediates $\cal F$ and $\cal W$ are defined by
\begin{eqnarray}
{\cal F}_{mi} =   \sum_n \sum_e
\widetilde{\langle m  n | i  e \rangle}\  t_n^e 
+ \sum_n \sum_{e,f}\langle m n | e f \rangle\ \widetilde{\tau_{i n}^{e f}},
\end{eqnarray}
\begin{eqnarray}
{\cal F}_{ae} =
 \sum_m \sum_f
\widetilde{\langle a m | e f \rangle}\  
t_m^f 
- \sum_{m,n} \sum_f 
\langle m n | e f\rangle\ \widetilde {\tau_{m n}^{a f}},
\end{eqnarray}
\begin{eqnarray}
{\cal F}_{me} =  \sum_n \sum_f \widetilde{\langle m n | e  f\rangle}\  t_{n}^{f},
\end{eqnarray}
\begin{eqnarray}
{\cal W}_{mnij} = \langle m n | i j\rangle +
P_{+}(mi,nj)\ \sum_e \langle m n | e j\rangle\ t_i^e 
+\sum_{e,f} \langle m n | e f \rangle \ \tau_{i j}^{e f},
\end{eqnarray}
\begin{eqnarray}
{\cal W}_{abef} = \langle a b | e f\rangle -
P_{+}(ae,bf)\ \sum_m
\langle m b | e  f \rangle\ t_m^a,
\end{eqnarray}
\begin{eqnarray}
{\cal W}_{mbej} =
{\langle mb | ej \rangle}
+ \sum_f {\langle mb | ef \rangle}\ t_j ^f 
- \sum_n {\langle mn|ej\rangle}\ t_b ^n 
+ \sum_n \sum_f \widetilde{\langle mn | ef\rangle}\ \left\{ \frac{1}{2} \widetilde{t_{jn}^{bf}}- t_j^f t_n^b \right\},
\end{eqnarray}
and
\begin{eqnarray}
{{\cal W}}_{mbje} =
\langle mb | je \rangle
+ \sum_f  \langle mb | fe \rangle\ t_j ^f
- \sum_n  \langle mn|je\rangle\ t_b ^n
- \sum_n \sum_f \langle mn | fe\rangle \ \left\{ \frac{1}{2} {t_{jn}^{fb}}+ t_j^f t_n^b \right\}.
\end{eqnarray}
and
\begin{eqnarray}
\widetilde{{\cal W}_{mbej}} = 2 {\cal W}_{mbej} + {\cal W}_{mbje}.
\end{eqnarray}
Finally, the symmetric permutation operator $P_+(pq,rs)$ is defined by
\begin{eqnarray}
P_+(pq,r) Z(pq,rs) = Z(pq,rs) + Z(qp,sr)
\end{eqnarray}
with $Z(pq,rs)$ as an arbitrary four-index quantity.
The calculations with {\sc QUENA} use orbital energies and two-electron integrals (in the MO representation) from a preceding {\sc CFOUR} calculation\cite{cfour,Matthews20c} which for the {\sc QUENA} computations in $C_{3v}$ were run in the largest Abelian subgroup of $C_{3v}$, namely $C_s$. This also means that the orbitals in the two-dimensional $E$ representation are chosen such that they transform either as $A'$ or $A''$. The results obtained with {\sc QUENA} have been validated such that the results (i.e., energies) agree for the $C_s$ and $C_{3v}$ computations carried out with {\sc QUENA} and also with those performed with {\sc CFOUR} using no symmetry or only $C_s$. 

\section{Exploratory Computations on NH$_3$ and PH$_3$}

To demonstrate the potential computational savings due to the exploitation of the full molecular point-group symmetry, we report in Tables~\ref{table1} and \ref{table2} the operation counts (of the ${\cal O}(M^5)$ and ${\cal O}(M^6)$ steps only) for CCSD computations on NH$_3$ (r(NH)=1.1 \AA , $\langle$(HNH) = 103.42$^\circ$ and PH$_3$ (r(PH)=1.42 \AA, $\langle$(HPH) = 119.63$^\circ$) when performed with $C_1$, $C_s$, and $C_{3v}$ symmetry using the cc-pVQZ basis set.\cite{Dunning89} 
The operation counts are given for the relevant terms (see previous section) individually together with the saving factor compared to the corresponding computations performed with lower symmetry. We refrain from reporting timings, as our pilot code has yet not been optimized. On the other side, the reported number of operations should provide information about potential savings that can be obtained with an optimal code, although the real savings will (due to the reduced size of the matrices in the matrix-matrix operations) remain somewhat smaller. 
\begin{landscape}
\begin{table}
\begin{tabular}{l|ccc}
term & $C_1$ & $C_s$ & $C_{3v}$ \ \\\hline\hline
\underline{formation of intermediates}\\
${\cal F}_{mi}$ & 2450000 & 843890 (2.9) & 188338 (13.0)    \\
${\cal F}_{ae}$ & 68600000 & 17878420 (3.8) & 3481788 (19.7) \\
${\cal W}_{mnij}$ & 12250000+87500 & 3513850+27515 (3.5) & 674094+5679 (18.1) \\
${\cal W}_{abef}$ & 1920800000 & 526886860 (3.6) & 90653844 (21.2) \\
${\cal W}_{mbej}\ {\rm and}\ {\cal W}_{mbje}$ & 2 x 343000000 + 2 x 68600000  & 2 x 88036900 +2 x 18959390 & 2 x 16544244  + 2 x 3294054  \\
 & + 2 x 2450000 & + 2 x 696485 (3.8) & + 2 x 125469 (20.7) \\
\underline{contractions in singles equations}\\
$t_2$ terms & 68600000 + 2450000 &  18959390 + 696485 (3.6) & 3391390 + 127585  (20.2) \\
\underline{contractions in doubles equations}\\
$\sum_m {\cal F}_{mi} t_{mj}^{ab}$ & 2450000 & 843890 (2.9) & 186222 (13.2)   \\
$\sum_e {\cal F}_{ae} t_{ij}^{eb}$ & 68600000 & 17878420 (3.8) & 3384452 (20.3) \\
$\sum_{m,n} {\cal W}_{mnij} {\tau}_{mn}^{ab}$ & 12250000 & 3513850 (3.5) & 674094 (18.2) \\
$\sum_{e,f}{\cal W}_{abef} {\tau}_{ij}^{ef}$ & 9604000000 & 2445152900 (3.9) & 431026844 (22.3) \\
$\sum_m \sum_e\widetilde {{\cal W}_{mbej}} \widetilde{t_{am}^{ie}}\ $   &&&\\
\hspace{0.5cm}${\rm and}\ \sum_m \sum_e{\cal W}_{mbje} t_{im}^{ea}$ 
& 2 x 343000000 & 2 x 88036900 (3.9) & 2 x 16544244 (20.7)\\

$t_1^2$ terms & 4 x  2450000 & 4 x 696485 (3.5) & 4 x 125469 (19.5) \\
$t_1$ terms &  68600000 +  2450000 &  18959390 + 696485 (3.6) & 3294054 + 125469 (20.8)\\
\hline
total & 13357487500 & 3450096635 (3.9)& 610731751 (21.9)\\
\end{tabular}
\caption{Number of operations 
with savings due to symmetry given in  parentheses for the ${\cal O}(M^5$) and ${\cal O}(M^6$) steps of a CCSD computation for NH$_3$ using the cc-pVQZ basis.}
\label{table1}
\end{table}
\end{landscape}
\begin{landscape}
\begin{table}
\begin{tabular}{l|ccc}
term & $C_1$ & $C_s$ & $C_{3v}$ \ \\\hline\hline
\underline{formation of intermediates}\\
${\cal F}_{mi}$ & 14288400 & 4729050 (3.0) & 1017462 (14.0)    \\
${\cal F}_{ae}$ & 222264000 &  57785140 (3.8) & 11308552 (19.7) \\
${\cal W}_{mnij}$ & 128595600+918540 & 35701450 + 277745 (3.6) &  6762758 + 55303 (19.0) \\
${\cal W}_{abef}$ & 3457440000& 942210476 (3.7) &  165114588 (20.9) \\
${\cal W}_{mbej}\ {\rm and}\ {\cal W}_{mbje}$ & 2 x 2000376000 + 2 x 14288400   & 2 x 511528500 + 2 x 60960290  & 2 x 95938688 + 2 x 10766682 \\
& + 2 x 222264000 & + 2 x 4007425 (3.9) & + 2 x 727133 (20.8) \\
\underline{contractions in singles equations}\\
$t_2$ terms & 222264000 + 14288400 & 60960290 + 4007425 (3.6) &  11156026 + 744061 (19.9)\\
\underline{contractions in doubles equations}\\
$\sum_m {\cal F}_{mi} t_{mj}^{ab}$ & 14288400 & 4729050 (3.0) & 1000534 (14.3)   \\
$\sum_e {\cal F}_{ae} t_{ij}^{eb}$ & 222264000 & 57785140 (3.8) & 10919208 (20.4) \\
$\sum_{m,n} {\cal W}_{mnij} {\tau}_{mn}^{ab}$ & 128595600 & 35701450 (3.6) & 6762758 (19.0) \\
$\sum_{e,f}{\cal W}_{abef} {\tau}_{ij}^{ef}$ &  31116960000 & 7902563684 (3.9) & 1417918180 (21.9) \\
$\sum_m \sum_e\widetilde {{\cal W}_{mbej}} \widetilde{t_{am}^{ie}} $  &&& \\
$ {\rm and}  \sum_m \sum_e{\cal W}_{mbje} t_{im}^{ea}$   & 2 x 2000376000 & 2 x 511528500 (3.9)& 2 x 95938688 (20.9)\\
$t_1^2$ terms & 4 x 14288400 & 4 x 4007425 (3.6) & 4 x 727133 (19.7)\\
$t_1$ terms & 222264000 + 14288400 & 60960290 + 4007425 (3.6) & 10766682 + 727133 (20.6)\\
\hline
total & 44310481740 &  11363497745 (3.9)& 2053904159 (21.6)\\
\end{tabular}
\caption{Number of operations 
with savings due to symmetry given in  parentheses for the ${\cal O}(M^5$) and ${\cal O}(M^6$) steps of a CCSD computation for PH$_3$ using the cc-pVQZ basis.}
\label{table2}
\end{table}
\end{landscape}
From the operation counts in both tables, we see that the calculations run with $C_{3v}$ symmetry require much fewer operations per iteration than the corresponding calculations run with $C_s$ or even $C_1$. While the savings of about 3.9 for both NH$_3$ and PH$_3$ in the calculations using $C_s$ compared to the calculation using no symmetry is in line with what was already observed in Ref.~\onlinecite{Stanton91a}, the calculation using $C_{3v}$ reduce the operation count by another factor of 5.5, resulting in overall savings compared to calculations using no symmetry of about 21.6 to 21.9. We note that the saving factors are the largest for the terms involving many virtual orbitals. Thus, while for the hole-hole ladder term the saving is only 18.2 for NH$_3$ (19.0 for PH$_3$), the savings are for the ring and particle-particle ladder term with 20.7 and 20.9 (ring terms) and 22.3 and 21.9 (particle-particle ladder) larger. The savings are very similar for NH$_3$ and PH$_3$ and similar results are also obtained for computations using the smaller cc-pVDZ and cc-pVTZ basis sets.\cite{Dunning89} The overall savings are here for all steps of a CCSD iteration: 19.5 (NH$_3$/cc-pVDZ),
21.1 (NH$_3$/cc-pVTZ), 
19.5 (PH$_3$/cc-pVDZ),
and 20.9 (PH$_3$/cc-PVTZ),
respectively. 
We thus conclude that exploitation of the full symmetry in cases with non-Abelian point-group symmetry can lead to additional savings (compared to treatments just using the largest Abelian subgroup) and that a general implementation would open the possibility of CC calculations on large and highly symmetric molecules (such as for example C$_{60}$ which so far are not possible with a standard CC code\footnote{CCSD calculation on C$_{60}$, however, are already possible using CC schemes that use Cholesky decomposition of two-electron integrals, see Ref.~\onlinecite{Nottoli23}.}).

We also like to comment on the maximal (or optimal) savings that in principle can be reached in the treatment of non-Abelian symmetry. Assuming an equal distribution of the orbitals among the different irreducible representations (thereby taking into account the degeneracy and attributing twice as many orbitals to the $E$ representation), one can expect savings of about 22.9 for the ${\cal O} (M^6)$ steps; we obtain in our examples savings of about 22.0. This is less than of the order of the square of the order of the group (which one obtains in the case of Abelian point groups and would be 36 for $C_{3v}$), but significant.

\section{Conclusions and Outlook}

In the present work, we demonstrate how the direct-product decomposition approach by Stanton {\it et al.}\cite{Stanton91a} for the treatment of Abelian point-group symmetry in CC calculations can be extended to non-Abelian point groups. The key findings are here that a block structure for higher-dimensional quantities such as the two-electron integrals and the amplitudes can be obtained by resolving the reducible products of two irreducible representations (in the case of $C_{3v}$ of the $E \otimes E$ product) into its irreducible representations. This also allows to eliminate redundancies (i.e., by just keeping one of the $E$ blocks) and enables to perform the ${\cal O}(M^6)$ contractions in the same way as in the Abelian case without a significant overhead for symmetry exploitation. However, complications arise (a) in the resorts of these higher-dimensional quantities, (b) when carrying out spin-adaptation, and (c) in case of the ${\cal O}(M^5)$ contractions. To overcome these complications we propose a strategy which employs both the reduced and the non-reduced representations of the involved higher-dimensional quantities, as all the
problematic steps are rather straightforward carried out in the non-reduced representation (apart from the fact that one has to deal with the ``missing integral/amplitude'' issue which is due to the fact that not all integrals in the calculation are available but when needed can be reconstructed from the available ones). It is not necessary to store both representations, as one can easily switch back and forth between them, a strategy which we use in our implementation within the new program package {\sc QUENA}. This program features a pilot implementation for the case of $C_{3v}$ symmetry and parts of the symmetry treatment so far are hard coded for that point group. However, we expect that generalization of our concepts to other non-Abelian point groups is straightforward.
Work along these lines is in progress, as also work concerned with the extension of the current concepts of symmetry exploitation to the perturbative treatment of triple excitations in the framework of CCSD(T).\cite{Raghavachari89} 


\section*{Acknowledgement}
The unexpected death of John F. Stanton in March 2025 was a great shock and a great loss for many of us. John was one of my (J.G.) best friends and with about 90 joint publications certainly by far my most important collaborator. Doing science with John was always fun. He was a great source of inspiration and he was always willing to take new challenges. The present work is in some way a reminiscence to our first joint paper (i.e., the one introducing the direct-product decomposition approach for CC calculations) and, thus, its dedication to his memory is more than appropriate. The authors thank Simon Blaschke and Jonas Greiner for discussions as well as Sophia Burger, Magdalena B\"uchner, and Stella Stopkowicz for helpful comments on the manuscript.

\bibliography{bibliography}

\end{document}